\documentstyle[stwol,axodraw]{article}
\input psfig
\begin{document}
\newcommand{\bfig}{\begin{center}\begin{picture}}
\newcommand{\efig}[1]{\end{picture}\\{\footnotesize #1}\end{center}}
\newcommand{\flin}[2]{\ArrowLine(#1)(#2)}
\newcommand{\wlin}[2]{\DashLine(#1)(#2){2.5}}
\newcommand{\zlin}[2]{\DashLine(#1)(#2){5}}
\newcommand{\glin}[3]{\Photon(#1)(#2){2}{#3}}
\newcommand{\lin}[2]{\Line(#1)(#2)}
\newcommand{\sof}{\SetOffset}
\newcommand{\bmip}[2]{\begin{minipage}[t]{#1pt}\bfig(#1,#2)}
\newcommand{\emip}[1]{\efig{#1}\end{minipage}}
\newcommand{\putk}[2]{\Text(#1)[r]{$p_{#2}$}}
\newcommand{\putp}[2]{\Text(#1)[l]{$p_{#2}$}}
\newcommand{\bq}{\begin{equation}}
\newcommand{\eq}{\end{equation}}
\newcommand{\bqa}{\begin{eqnarray}}
\newcommand{\eqa}{\end{eqnarray}}
\newcommand{\nl}{\nonumber \\}
\newcommand{\eqn}[1]{eq. (\ref{#1})}
\newcommand{\eqs}[1]{eqs. (\ref{#1})}
\newcommand{\ibidem}{{\it ibidem\/},}
\newcommand{\vpb}{}
\newcommand{\p}[1]{{\scriptstyle{\,(#1)}}}
\newcommand{\gev}{\mbox{GeV}}
\newcommand{\mev}{\mbox{MeV}}

\title{
\vskip -30pt
\begin{flushleft}
{\footnotesize {\em Talk given  at XXVIII Int. Conference on High Energy
Physics, Warsaw, 25-31 July 1996.~~~~~~~~~~~~~~~~~~}PSI-PR-96-20}
\end{flushleft}
\vskip 20pt
LEP2 PHYSICS AND EVENT GENERATORS}
\author{ R. PITTAU }

\address{ Paul Scherrer Institut, CH-5232 Villigen PSI, Switzerland}

\twocolumn[\maketitle\abstracts{
The present status of four-fermion calculations and event generators
for LEP2 physics is reviewed. Perspectives for future improvements are
given.}]

\section{Introduction}
At LEP2, both {\em Standard} and {\em New} physics can be studied \cite{rep}.
On one hand, precision tests of the Standard Model are possible, first of
all by measuring the $W$ mass, but also by studying two-fermion
processes off-resonance, $\gamma \gamma$ physics and $QCD$.
On the other hand, the possibility of new physics discoveries exists.
For example, searches for supersymmetric and/or new particles can be
performed and the trilinear gauge boson vertex investigated in detail.
Higgs physics lies somehow between those two categories:
in case of discovery at LEP2, it could be hard to decide on the
standard or non-standard nature of a neutral Higgs.

Due to the small cross sections, the number of collected events at LEP2
will be limited and two different attitudes can be adopted. One could
think that, because of the low statistics, precise theoretical
calculations are unimportant, but also -on the contrary- that, just
because of the limited data, accurate theoretical
knowledge is necessary in order to reduce the systematic error and 
extract as much information as possible. 
The choice between those two strategies is not matter of taste, but
depends on the type of physics one is interested in. 
In fact, for discovery physics, one does not need very
sophisticated tools. On the contrary, performing precision physics at
LEP2 requires a dedicated effort. Forgetting that can easily lead to
an bad underestimation of the systematic error in the precision measurements.

\noindent At LEP1 the enormous statistics allowed a strong interplay
theory-experiment, which is not possible at LEP2. Therefore,
at least for precision physics, theory must take over. As a consequence,
when performing precision measurements, LEP2 Event Generators must be 
dedicated codes including loop corrections and all kind of backgrounds,
while, for discovery physics, tree level signal programs
are in general sufficient, unless the effects induced by new physics are
expected to be tiny \footnote{For example, a consistent study of 
the anomalous couplings has necessarily to be performed at the level of a 
four-fermion Event Generator.}.

In the following, I shall concentrate on four-fermion physics
in $e^+ e^-$ collisions, by reviewing the present knowledge on the topic.
I shall analyze the various contributions, pointing out
what is still missing and should be computed for LEP2 experiments.
\section{Four-fermion physics and codes}
\subsection{Tree level}
Calculations involving four fermions in the final state are unavoidable at
LEP2. In $M_W$ measurement the relevant process is the $W^+ W^-$
production mechanism of fig. 1, but, since $\Gamma_W \ne 0$, 
the actual measured signal is a four-fermion final state.
Therefore, one is led to consider decaying $W$'s together with all
contributing four-fermion background diagrams.
\bfig(100,125)
\SetScale{1}
\SetWidth{1}
\sof(-81,10)
\flin{40,0}{65,25} \flin{65,25}{65,70} \flin{65,70}{40,95}
\wlin{65,25}{105,25}
\wlin{65,70}{105,70}
\Text(38,0)[tr]{$e^-$}
\Text(38,95)[br]{$e^+$}
\Text(87,30)[b]{$W$}
\Text(87,75)[b]{$W$}
\sof(12,32)
\flin{40,0}{65,25} \flin{65,25}{40,50}
\Text(38,0)[tr]{$e^-$}
\Text(38,50)[br]{$e^+$}
\wlin{65,25}{107,25}
\wlin{107,25}{127,45}
\wlin{107,25}{127,5}
\Text(130,50)[bl]{$W$}
\Text(130,0)[tl]{$W$}
\Text(85,31)[b]{Z, $\gamma$}
\efig{Figure 1: $W^+ W^-$ signal diagrams.}

\vspace{6pt}

\noindent Analogously, a four-fermion final state is the measured signal
for Higgs physics and trilinear anomalous couplings studies 
(see fig. 2). Disregarding fermion masses, there are in total 27 leptonic
four-fermion final states, 42 semileptonic processes
and 17 hadronic channels \cite{ber}.
\bfig(100,175)
\SetScale{1}
\sof(-30,100)
\Text(-20,60)[l]{\bf (a)}
\flin{40,0}{65,25} \flin{65,25}{40,50}
\Text(38,0)[tr]{$e^-$}
\Text(38,50)[br]{$e^+$}
\wlin{65,25}{107,25}
\wlin{107,25}{127,45}
\wlin{107,25}{127,5}
\Text(130,50)[bl]{$Z$}
\Text(130,0)[tl]{$H$}
\Text(87.5,31)[b]{$Z$}
\Text(-20,-17)[l]{\bf (b)}
\sof(-75,13)
\flin{40,0}{65,25}\flin{65,25}{40,50}
\Text(38,0)[tr]{$e^-$}
\Text(38,50)[br]{$e^+$}
\wlin{65,25}{95,25}
\wlin{95,25}{120,50}
\wlin{95,25}{120,0}
\Text(124,50)[bl]{$W$}
\Text(124,0)[tl]{$W$}
\Text(77,32)[b]{$Z$, $\gamma$}
\GOval(95,25)(5,5)(0){0.65}
\sof(35,13)
\flin{100,50}{70,50}\flin{70,50}{40,50}
\flin{40,0}{70,0}\flin{70,0}{100,0} 
\Text(32,4)[bl]{$e^-$}
\Text(32,46)[tl]{$e^+$}
\Text(105,50)[l]{$\bar{\nu_e}$}
\wlin{70,50}{85,25}
\wlin{70,0}{85,25}
\wlin{85,25}{115,25}
\GOval(85,25)(5,5)(0){0.65}
\Text(116,31)[r]{$W$}
\Text(73,36)[r]{$W$}
\Text(73,12)[r]{$Z$, $\gamma$}
\Text(108,4)[bl]{$e^-$}
\efig{Figure 2: Higgs signal (a) and trilinear couplings (b).}

\vspace{10pt}

\noindent The available four-fermion tree level codes are listed in table 1.  

\footnotesize

\begin{table}[h]
\begin{center}
\begin{tabular}{|c|c|c|c|c|c|} \hline
 Program   & type & Diagrams & AC & $m_f$& Higgs\\
\hline
\hline
{\tt Alpha}     &MC&all      &$-$&+&$-$\\
\hline
{\tt Comphep}   &EG&all      &$-$&+&+\\
\hline
{\tt Erato}     &MC&cc11/cc20 &+&$-$&$-$\\
\hline
{\tt Excalibur} &MC&all      &+&$-$&$-$\\
\hline
{\tt Gentle}    &SA&cc11/nc32&$-$&$\pm$&+\\
\hline
{\tt Grc4f}     &EG&all      &+&+&$-$\\
\hline
{\tt Higgspv}   &EG&nnc      &$-$&$\pm$&+\\
\hline
{\tt Koralw}    &EG&cc11     &+&$\pm$&$-$\\
\hline
{\tt Lepww}     &EG&cc03     &+&$-$&$-$\\
\hline
{\tt Lepww02}   &EG&cc03     &$-$&$\pm$&$-$\\
\hline
{\tt Pythia}    &EG&cc03     &$-$&$\pm$&+\\
\hline
{\tt Wopper}    &EG&cc03     &$-$&$\pm$&$-$\\
\hline
{\tt Wphact}    &MC&all      &+&+&+\\
\hline
{\tt Wto}       &Int&ncc    &$-$&$-$&+\\
\hline
{\tt Wwf}       &EG&cc11     &+&+&$-$\\
\hline
{\tt Wwgenpv}   &EG&cc11/cc20&$-$&$\pm$&$-$\\
\hline
{\tt Hzha}      &EG&Susy signal&$-$&$+$&+\\
\hline 
\end{tabular}
\end{center}
\caption[.]{Available four-fermion programs (MC= Monte Carlo, EG=
Event Generator, SA= Semi-Analytic, Int= Integrator).
The included diagrams are given in column 3 using the classification
of ref. (\,\cite{BLR}). In columns 4, 5 and 6 a $+$ ($-$) sign is written if the
program includes (does not include) anomalous couplings, $m_f \ne 0$
and Higgs diagrams. $\pm$ denotes approximate treatment of the fermion masses.}
\end{table}

\normalsize
 
\noindent Neglecting fermion masses is a good
approximation at LEP2 energies except for Higgs production (couplings 
$\propto m_f$) and studies involving electrons in the very forward
region (t-channel photon diagrams become singular in the limit $m_e \to  0$).  

\noindent Detailed comparison among codes can be found 
in ref. (\,\cite{rep1}). 
In fig. 3, I show the typical result of a tuned comparison among
dedicated programs, namely codes including both signal and background
diagrams.

\begin{center} 
\begin{picture}(160,150)
\SetScale{0.5}
\SetOffset(25,-13)
\Text(55,155)[b]{{-1}}
\Text(70,155)[b]  {{0}}
\Text(85,155)[b]{{1}}
\Text(112.5,155)[b]{{(permill)}}
\Text(-30,135)[l]{{\tt Comphep}}
\Text(-30,120)[l]{{\tt Erato}}
\Text(-30,105)[l]{{\tt Excalibur}}
\Text(-30,90)[l]{{\tt Grc4f}}
\Text(-30,75)[l]{{\tt Wphact}}
\Text(-30,60)[l]{{\tt Wto}}
\Text(-30,45)[l]{{\tt Wwgenpv}}
\Text(70,26)[t]  {{.614} pb}
  \LinAxis(17.2,300)(262.8,300)(8,1,-5,0,1.5)
  \LinAxis(17.2,60)(262.8,60)(2,1,5,0,1.5)
\SetWidth{5}
  \Line(101,270)(161,270) 
  \Line(124,240)(154,240)
  \Line(130,210)(150,210) 
  \Line(125,180)(145,180)
  \Line(120,150)(150,150)
  \Line(122,120)(152,120)
  \Line(104,90)(164,90)   
\efig{Figure 3: $\sigma(e^- \bar \nu_e u \bar d)$ at $\sqrt{s}= 190~GeV$.
ADLO/TH cuts as in ref. (\,\cite{rep1}), ISR included.}

\vspace{10pt}

\noindent A last comment is in order. 
Giving by hand a width to the bosons  
in tree level calculations breaks gauge invariance.
A solution to this is the fermion loop (FL) approach of ref.
(\,\cite{pass9}), in which the imaginary part of the relevant 
one-loop diagrams 
is included to restore gauge independence. The deviation 
among the naive running width prescription, the FL result and the fixed
width approach (constant complex masses in all propagators) is given
in figure 4. 
One convinces oneself that the fixed
width scheme - although without theoretical justification - 
numerically agrees with the FL result.
\subsection{Electroweak radiative corrections}
At LEP2, radiative corrections turn out to be extremely important for
precision physics. In $M_W$ measurement, the expected shift in the 
reconstructed mass is \cite{katsa} $\Delta M_W =~ <E_\gamma> M_W/\sqrt{s}$ 
where $<E_\gamma>$ is the average energy lost by QED radiation.
To have control on $<E_\gamma>$ requires, in principle, an evaluation
of the one-loop QED corrections to $e^+ e^- \to$ 4 fermions, namely, 
computing objects like the six-point diagram in fig
5. Furthermore, unlike at LEP1, QED and weak corrections are not
separately gauge invariant. Therefore, for the sake of consistency,
one should also include the full set of one-loop weak corrections. 
\begin{figure}
\center
\vskip -4.1cm
\hskip -0.85cm
\psfig{figure=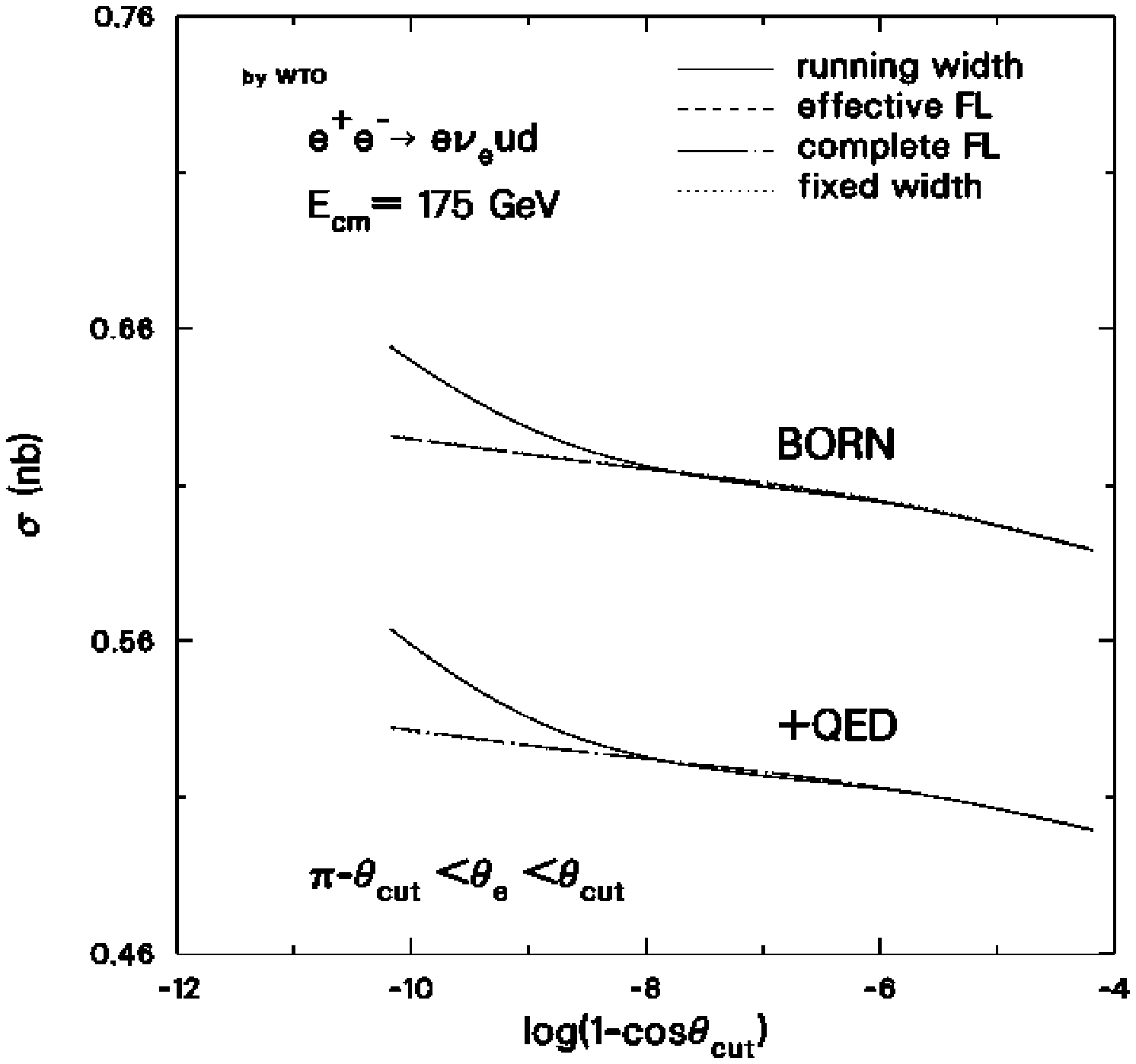,height= 4.85in}
\vskip -0.88cm
{\footnotesize Figure 4: $\sigma(e^- \bar \nu_e u \bar d)$ 
as a function of the angular cut of the electron with different
schemes for the widths.}
\vskip -0.5cm
\end{figure}

\noindent At the moment, what is available is the implementation of the
resummed (universal) leading log (LL) part of the initial state QED radiation
in all programs of table 1. Some of them can include LL final state
radiation and models for the generation of 1 photon with finite $p_T$.
One code ({\tt Gentle}) also computes part of the subleading (non
universal) QED corrections, by using the splitting techniques of ref.
(\,\cite{split}).
\bfig(160,116)
\SetScale{1}
\sof(-17,10)
\flin{40,0}{65,25} \flin{65,25}{65,70} \flin{65,70}{40,95}
\Text(37,0)[tr]{$e^-$}
\Text(37,95)[br]{$e^+$}
\wlin{65,25}{110,25}
\wlin{65,70}{110,70}
\flin{145,65}{110,70}\flin{110,70}{135,95}
\flin{135,0}{110,25}\flin{110,25}{145,30}
\glin{135,29}{135,66}{8}
 \Text(139,47)[l]{$\gamma$}
 \Text(87,29)[b]{W}
 \Text(87,74)[b]{W}
\efig{Figure 5: One-loop LEP2 QED diagram.}

\vspace{10pt}

\noindent As for the pure weak corrections, full one-loop
calculations are available for on-shell $W$'s only \cite{fred} and for the
factorizable weak part in off-shell $W^+W^-$ production \cite{geert}. 
At present, a full four-fermion electroweak calculation seems to be
out of reach (although some promising techniques have been recently 
introduced \cite{multiloop}). A gauge invariant and modular approach
to the problem is the pole scheme described in ref. (\,\cite{rep2}).
\subsection{QCD contributions and loop corrections}
QCD enters the game in two different ways. Firstly, diagrams
like those in fig. 6 contribute to four-quarks or four-jets final
states as a background (for example) in $M_W$ reconstruction \cite{4quark}. 

\noindent All dedicated codes in table 1 can easily
include them, when computing four-quark processes.
\bfig(160,102)
\sof(-50,5)
\flin{45,5}{65,25}\flin{65,25}{45,45}
\Text(42,5)[rt]{$e^-$}
\Text(42,45)[rb]{$e^+$}
\wlin{65,25}{85,25}
\Text(75,30)[b]{$Z, \gamma$}
\sof(-65,5)
\flin{120,5}{100,25} \flin{100,25}{120,45}
\Gluon(113.5,38.5)(115,65){2}{4}\flin{135,45}{115,65}\flin{115,65}{135,85}
\Text(105,55)[t]{$g$}
\sof(60,5)
\flin{45,5}{65,25}\flin{65,25}{45,45}
\Text(42,5)[rt]{$e^-$}
\Text(42,45)[rb]{$e^+$}
\wlin{65,25}{85,25}
\Text(75,30)[b]{$Z, \gamma$}
\sof(45,5)
\flin{120,5}{100,25}\flin{100,25}{120,45}
\Gluon(113,38)(131,38){2}{3}
\Gluon(113,12)(131,12){2}{3}
\Text(136,38)[l]{$g$}
\Text(136,12)[l]{$g$}
\efig{Figure 6: Some QCD diagrams in four-jets production.}

\vspace{10pt}

\noindent Secondly, radiative QCD corrections are present.
Typical QCD loop diagrams for semi-leptonic final states, 
are shown in fig. 7. They have to be considered together with real
gluonic emission to give the physical (infrared safe) cross section 
in 2 leptons + 2 jets.
\bfig(160,102)
\sof(-50,5)
\flin{45,5}{65,25}\flin{65,25}{45,45}
\Text(42,5)[rt]{$e^-$}
\Text(42,45)[rb]{$e^+$}
\wlin{65,25}{85,25}
\Text(75,30)[b]{$Z, \gamma$}
\sof(-65,5)
\flin{120,5}{100,25} \flin{100,25}{120,45}
\wlin{113.5,38.5}{115,65}\flin{135,45}{115,65}\flin{115,65}{135,85}
\Gluon(132,82)(132,48){2}{5}
\Text(140,85)[lb]{$q$}
\Text(140,45)[lt]{${\bar q}^\prime$}
\Text(140,65)[lt]{$g$}
\sof(60,5)
\flin{45,5}{65,25}\flin{65,25}{45,45}
\Text(42,5)[rt]{$e^-$}
\Text(42,45)[rb]{$e^+$}
\wlin{65,25}{85,25}
\Text(75,30)[b]{$Z, \gamma$}
\sof(45,5)
\flin{120,5}{100,25}\flin{100,25}{120,45}
\Gluon(117,42)(117,8){2}{5}
\Text(125,45)[lt]{$q$}
\Text(125,5)[lt]{${\bar q}^\prime$}
\Text(125,25)[lt]{$g$}
\wlin{113.5,38.5}{115,65}\flin{135,45}{115,65}\flin{115,65}{135,85}
\efig{Figure 7: QCD virtual diagrams in semileptonic channels.}

\vspace{10pt}

\noindent Such radiative QCD corrections can be naively taken into account,
for semileptonic processes at LEP2 energies, by rescaling $W$ width and
cross section as follows \cite{rep1}
\bqa
\Gamma_W \to \Gamma_W\,(1+\frac{2}{3}\,\frac{\alpha_s}{\pi})\,,~~~~
\sigma \to \sigma\,(1+\frac{\alpha_s}{\pi})\,.
\eqa
Strictly speaking, the above replacements give the correct result for $W^+W^-$
production diagrams only, without cuts. Recently, an 
exact QCD one-loop calculation has been worked out for the
channel $\mu^- \bar \nu_\mu u\bar d$ \cite{qcd}. With ADLO/TH cuts \cite{rep1}
a good agreement between the naive QCD approach and the exact
calculation has been found, except for angular distributions (see fig. 8).
\begin{figure}
\center
\vskip -0.9 cm
\hskip 0.6cm 
\psfig{figure=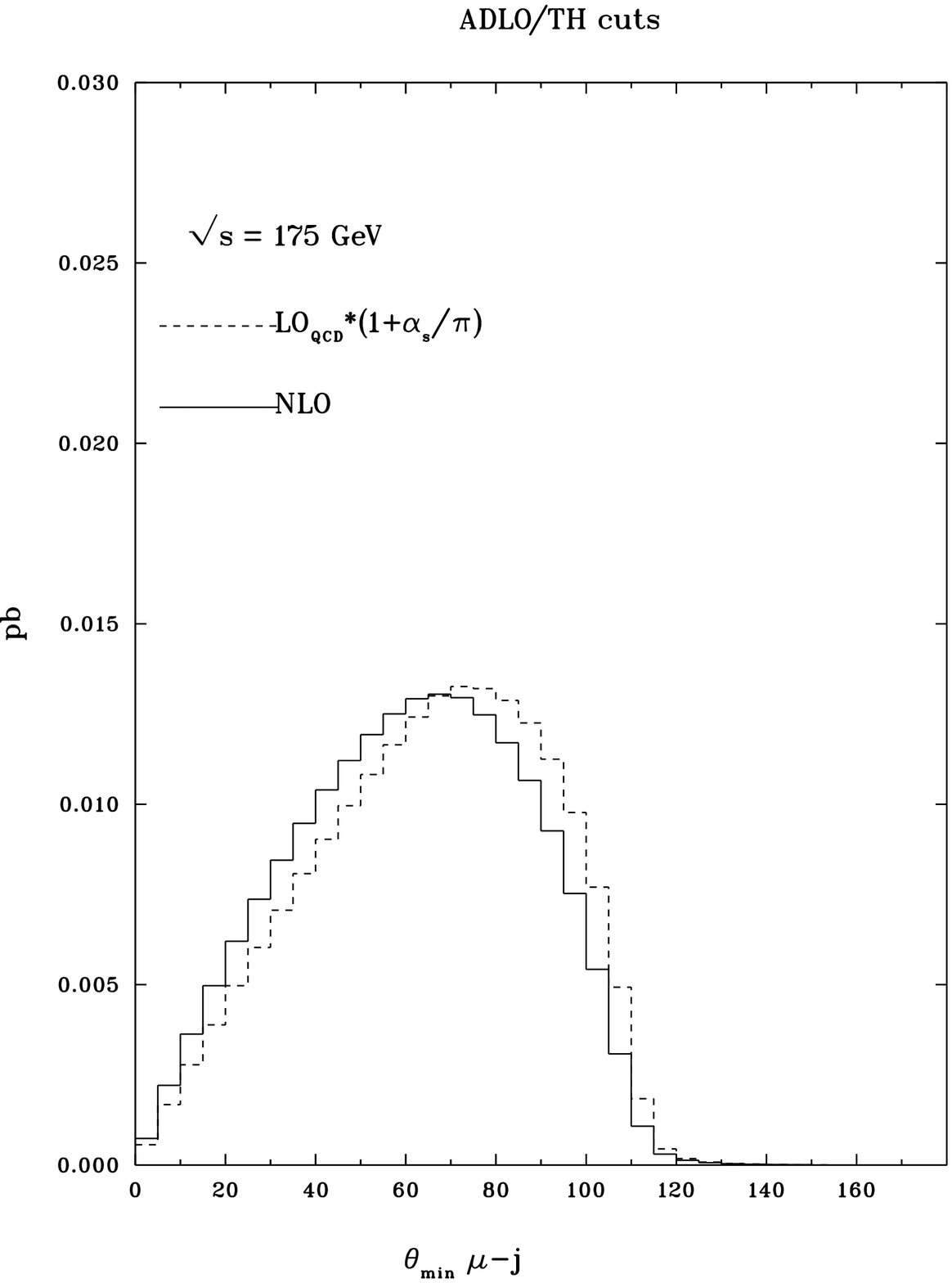,height= 4.4in}
\vskip -0.6 cm
{\footnotesize Figure 8: Distribution of the minimum angle between 
$\mu^-$ and jet in $\mu^- \bar \nu_\mu u\bar d$ production, using naive
(dashed) and exact (solid) one-loop QCD radiative corrections.}
\vskip -0.4 cm
\end{figure}

\noindent When dealing with QCD one also faces non perturbative phenomena. Most
of the knowledge on hadronization, collected at LEP1, can be directly
translated to LEP2 physics, with the exception of color reconnection and
Bose-Eistein effects in four-jet production \cite{rep2}, for which information will
have to be extracted from the LEP2 data. That may require a deep knowledge
of the perturbative QCD contributions in order to disentangle the
non-perturbative part. While all ingredients for computing ${\cal O}
(\alpha_s)$ loop corrections to four-jet production via electroweak
interactions are already available \cite{qcd}, a calculation of the pure
gluonic part (namely loop corrections to diagrams in fig. 6) is still missing.

\section{Conclusions}
Tree level four-fermion physics is in good shape. All processes
can be computed including, where necessary, fermion masses and a gauge
invariant solution exists for dealing with unstable particles. 

\noindent The available codes
have been successfully cross-checked, reaching high  
technical precision. However, the latter does not imply small
theoretical errors. Reducing theoretical uncertainties means
incorporating new contributions in the calculations, namely including
the loop corrections.
While progress has been recently made in QCD (at least for
semileptonic processes), our knowledge of the electroweak loop
corrections in four-fermion production is, at present, at the LL level only. 

\noindent A deeper understanding of the electroweak loop effects has
to be reached, expecially to meet the task $\Delta M_W = 50~ MeV$.
A first step in that direction could be employing the techniques in
ref. (\,\cite{multiloop}) to compute the photonic loop corrections 
in off-shell $W^+W^-$ production,
that are anyway one of the basic ingredients of the full calculation. 
A different contribution will be soon provided by the authors of ref.
(\,\cite{pass9}), that are working out the fermionic set of loop corrections for
$e^-\bar  \nu_e u \bar d$. 

\noindent Joining the forces of all peoples working on
radiative corrections in four-fermion physics, will soon become 
desirable to overcome the technical difficulties and reach
a satisfactory understanding of this very complicated subject.
\section*{Acknowledgment}
I wish to thank G. Passarino for supplying me with Figure 4 and for
useful discussions.

\end{document}